\documentclass[twocolumn,aps,amsmath,amssymb,floatfix,showpacs,superscriptaddress]{revtex4}
\usepackage{graphicx}
\begin{document}
\title{Scaling Behavior of Threshold Epidemics} 
\author{E.~Ben-Naim}
\affiliation{Theoretical
Division and Center for Nonlinear Studies, Los Alamos National
Laboratory, Los Alamos, NM 87545, USA}
\author{P.~L.~Krapivsky}
\affiliation{Department of Physics, Boston University, Boston, MA 02215, USA}
\begin{abstract}
We study the classic Susceptible-Infected-Recovered (SIR) model for
the spread of an infectious disease.  In this stochastic process,
there are two competing mechanism: infection and recovery. Susceptible
individuals may contract the disease from infected individuals, while
infected ones recover from the disease at a constant rate and are
never infected again.  Our focus is the behavior at the epidemic
threshold where the rates of the infection and recovery processes
balance.  In the infinite population limit, we establish analytically
scaling rules for the time-dependent distribution functions that
characterize the sizes of the infected and the recovered
sub-populations.  Using heuristic arguments, we also obtain scaling
laws for the size and duration of the epidemic outbreaks as a function
of the total population.  We perform numerical simulations to verify
the scaling predictions and discuss the consequences of these scaling
laws for near-threshold epidemic outbreaks.
\end{abstract}
\maketitle

\section{Introduction}

The study of epidemics has a long and fascinating history that dates
back to Daniel Bernoulli who modeled the spread of smallpox
\cite{McN,O,Ber}. Theoretical models have been quite successful in
describing the spread of infectious diseases. Closely related models
have been applied to a truly remarkable set of contagious processes
including HIV \cite{per,nm}, computer viruses \cite{lud}, spread of
technological innovations \cite{R03,KRV11}, outbreaks of
social and political unrest \cite{G78,L94}, and rumor propagation
\cite{dk}.

The convenient deterministic framework is most commonly used to model
the spread of an epidemic.  In this formulation, coupled ordinary
differential equations describe the evolution of macroscopic
properties such as the average number of infected, the average number
of recovered, and so on \cite{jdm}. However, the deterministic
approach cannot capture fluctuations which are inevitable due to the
finite size of the population. In many scenarios such as the spread of
an infection in an isolated boarding school with say, a hundred
children, statistical fluctuations are certainly significant.  The
stochastic framework which involves evolution equations for the entire
probability distribution is instead required to describe finite
populations \cite{McK,Bp,ntjb,MT,AB,book}. Theoretical analysis of the
stochastic framework is challenging, even for the most basic infection
processes \cite{ML,bk,KS,ML2,Dutch,KS1}, and many questions concerning
finite-size effects and extremal properties of the probability
distribution functions remain unanswered.

In this paper, we investigate the stochastic version of the classic
Susceptible-Infected-Recovered (SIR) infection process.  In this
model, the population consists of susceptible, infected, and recovered
individuals, and the infection spreads through contact between
infected and susceptible members of the community
\cite{ntjb,ma,hwh,jdm}.  Each infected individual spreads the
infection at a certain rate, denoted by $\alpha$, while infected
individuals recover at a rate set to one.  The epidemic threshold is
$\alpha=1$. Below the threshold, the infection cannot maintain itself;
above the threshold, the infection can take off.

We focus on epidemic outbreaks at or near the threshold
\cite{hwh,wd,hed,lp,bc,ad,jg}.  Such ``threshold epidemics'' are
especially interesting. From a theoretical viewpoint, the infection
and the recovery processes are in some sense in perfect balance
precisely at the threshold. While most infections are small, large
outbreaks may occasionally happen, and hence, threshold epidemics
exhibit large fluctuations.  From a practical viewpoint, human efforts
at disease prevention reduce the infection rate thereby driving
infectious diseases from the pandemic to the endemic phase
\cite{ma}. Meanwhile, natural evolution may increase the infection
rate of diseases hovering just below the threshold, thereby enhancing
the likelihood of an outbreak \cite{antia}. These epidemics are
subtle: they may be difficult to detect as well as difficult to
control.

We first consider infinite populations and focus on the behavior
precisely at the epidemic threshold.  We analyze the rate equation for
the probability $P_{i,r}(t)$ that the number of infected equals $i$
and the number of recovered equals $r$ at time $t$. The typical number
of infected grows linearly with time while the typical number of
recovered grows quadratically with time. We show that the joint
distribution function obeys the scaling form
\begin{equation}
\label{Pir-scal}
P_{i,r}(t)\simeq t^{-4}\,\Phi\left(i\,t^{-1},r\,t^{-2}\right).
\end{equation}
We obtain the Laplace transform of the scaling function
$\Phi(\xi,\eta)$ analytically, and present a comprehensive asymptotic
analysis. We also discuss scaling properties of the respective
single-population distribution functions for the number of infected or
the number of removed.

Next, we consider finite populations.  We combine the infinite
population results with heuristic arguments to derive scaling laws for
the size and duration of outbreaks in a finite population of size $N$.
At or near the epidemic threshold, the effective infection rate is
reduced because the population is finite.  We find that the maximal
size of the outbreak, $M$, and the maximal duration of the outbreak,
$T$, obey the nontrivial scaling laws
\begin{equation}
\label{MT}
M\sim N^{2/3},\quad{\rm and}\quad T\sim N^{1/3}.
\end{equation}
We also formulate the range of infection rates near the epidemic
threshold for which these scaling laws apply.

The rest of this paper is organized as follows.  We begin with the
definition of the infection process, given in Section II. In section
III, we treat the infinite population limit by using the rate equation
approach. First, we derive the outbreak size distribution.  We then
compute two single-variable distributions --- the probability
distribution for the number of infected individuals and the
probability distribution of the number of recovered individuals.
Next, we discuss scaling and extremal properties of the joint
distribution function. In section IV, we consider finite populations
and obtain finite-size scaling laws for the size and duration of the
outbreaks. We discuss the results in section V, and the appendix
details some necessary inverse Laplace transforms. 

\section{The infection process}

The spread of infectious diseases has been widely studied on regular
and irregular spatial structures such as lattices
\cite{pg,sa,ad1,hin,tibor,dickman,jdmII,wss,ziff} and
complex networks \cite{pv,ml,mejn}.  Nevertheless, many
characteristics of stochastic epidemics, e.g.
finite-size properties, remain open questions even for perfectly mixed
populations. Throughout this paper we assume that the populations are
well-mixed so that any infected individual can spread the disease to
any susceptible individuals.

In the ubiquitous Susceptible-Infected-Recovered (SIR) infection
process, the population includes three types of of individuals:
\begin{itemize}
\item[{\bf S}] Uninfected individuals who are susceptible to the infection.
\item[{\bf I}] Infected individuals who are actively spreading the
disease.
\item[{\bf R}] Individuals who are neither infected nor susceptible
  including those who have been infected and subsequently recovered,
  or became immune, or removed.
\end{itemize}
An individual may proceed from type ${\bf S}$ to ${\bf I}$ to ${\bf
  R}$. In this simplified model, all infected individuals may spread
the disease.

We investigate the continuous-time version of SIR infection process
\cite{ntjb}.  At any given moment, the population consists of $s$
susceptible, $i$ infected, and $r$ recovered individuals; the size of
the total population, $N=s+i+r$, remains fixed.  The sub-populations
change due to two competing processes --- infection and recovery.  An
infected individual may infect a susceptible one with rate $\alpha/N$,
where $\alpha$ is the infection rate:
\begin{equation}
\label{infection}
(s,i,r)\buildrel {\alpha si}/N \over \longrightarrow (s-1,i+1,r).
\end{equation}
The overall infection rate is proportional to the size of the
susceptible population times the size of the infected population.
Infected individuals recover at a constant rate,
\begin{equation}
\label{recovery}
(s,i,r)\buildrel i \over \longrightarrow (s,i-1,r+1).
\end{equation}
Without loss of generality, the recovery rate is set to one. 

We consider the natural initial condition where a single infected
individual is embedded in a ``sea'' of susceptible individuals, that
is, $(s,i,r)=(N-1,1,0)$ at time $t=0$. The infection process ends when
no infected individuals remain, $(s,i,r)=(N-n,0,n)$. The final number
infected, $n$, measures the size of the epidemic outbreak.

\section{Infinite Populations}

Since the infection process is random, the size of the epidemic
outbreak is a stochastic quantity. First, we investigate the
distribution of this quantity as its basic characteristics show how
the infection process can be in one of two phases: an endemic phase
where a microscopic number of individuals are infected and a pandemic
phase where a macroscopic number of individuals are infected.

In this section we analyze the infinite population limit, $N\to
\infty$.  Let $G_n$ be the probability that the size of the epidemic
outbreak equals $n$.  For the initial condition $(s,i,r)=(N-1,1,0)$,
the total infection rate and the total recovery rate are $\alpha$ and
$1$ as follows from \eqref{infection}--\eqref{recovery}.  In the very
first step, the infected individual either recovers, or a second
individual become infected.  The former happens with probability
$G_1=1/(1+\alpha)$, while the latter occurs with the complementary
probability $\alpha/(1+\alpha)$.  Hence, the distribution $G_n$ obeys
the recursion equation
\begin{equation}
\label{Gn-eq}
G_n=\frac{\alpha}{1+\alpha}
\sum_{i+j=n}G_iG_j+\frac{1}{1+\alpha}\,\delta_{n,1}\, ,
\end{equation}
valid for all $n\geq 1$.  The convolution term corresponds to the case
where an additional individual become infected --- in this situation
there are two infection processes, and in the infinite population
limit these two processes are completely independent
\cite{teh,branch}.  Starting with $G_1=1/(1+\alpha)$, the recursion
equation gives $G_2=\alpha/(1+\alpha)^3$, $G_3=2\alpha/(1+\alpha)^5$,
etc.

We now introduce the generating function
\begin{equation}
\label{Gz-def}
{\cal G}(z)=\sum_{n\geq 1} G_nz^n.
\end{equation}
Using the generating function we convert the infinite set of recursion
equations (\ref{Gn-eq}) into the quadratic equation,
$(1+\alpha){\cal G}=\alpha\,{\cal G}^2+z$.  Out of the two solutions, 
the proper one satisfies the requirement $\mathcal{G}(z=0)=0$, 
\begin{equation}
\label{Gz}
{\cal G}(z)=\frac{1+\alpha-\sqrt{(1+\alpha)^2-4\alpha z}}{2\alpha}\,.
\end{equation}

The behavior of $G(z)$ in the limit $z\to 1$ reveals the basic 
characteristics of the infection process.  The probability $G_{\rm
  finite}$ that only a {\em finite} number of individuals are infected
is given by
\begin{equation}
\label{G-finite}
G_{\rm finite}=
\begin{cases}
{\displaystyle 1}\qquad                      & \alpha\leq 1;\cr
{\displaystyle \alpha^{-1}}\qquad       & \alpha>1;\cr
\end{cases}
\end{equation}
The derivation of \eqref{G-finite} follows from \eqref{Gz} and
\hbox{$G_{\rm finite}=\sum_{n\geq 1} G_n={\cal G}(1)$}.  The threshold
value $\alpha_c=1$ separates two regimes of behavior --- an endemic
regime and a pandemic regime. Below the threshold, the number of
infected individuals is always finite. Above the threshold, there is a
finite probability that a finite {\em fraction} of individuals becomes
infected, and as a consequence, the average size of the epidemic
outbreak is macroscopic, that is, proportional to the size of the
total population.

Also, the average number of infected individuals readily follows from
the generating function, viz. \hbox{$\langle n\rangle=\sum_n
nG_n={\cal G}'(1)$}. Equation \eqref{Gz} gives
\begin{equation}
\label{nav}
\langle n\rangle=\frac{1}{1-\alpha}, \qquad \alpha<1.
\end{equation}
As expected, the size of the outbreak diverges in the vicinity of the
threshold.

To find $G_n$, we expand the generating function $G(z)$ in powers of
$z$.  The distribution of outbreak size is a product of an exponential
and an algebraic factor, expressed as a ratio of Gamma functions,
\begin{equation*}
G_n=\frac{1}{(1+\alpha)\,\sqrt{\pi}}\,
\frac{\Gamma\left(n-\frac{1}{2}\right)}
{\Gamma(n+1)}
\left[1-\left(\frac{1-\alpha}{1+\alpha}\right)^2\right]^{n-1}.
\end{equation*}
By using the asymptotic behavior $\Gamma(x+a)/\Gamma(x)\simeq x^a$ as
$x\to\infty$, we conclude that at the epidemic threshold, $\alpha=1$,
the size distribution has a power-law tail
\begin{equation}
\label{Gn-large}
G_n\simeq (4\pi)^{-1/2}\,n^{-3/2}, 
\end{equation}
as $n\to \infty$.  For threshold epidemics, there is balance between
infection and recovery, as indicated by
\eqref{infection}--\eqref{recovery}. While the majority of outbreaks
are small, the algebraic behavior \eqref{Gn-large} suggests that there
is a considerable likelihood for large outbreaks to occur.

In the remainder of this section, we focus on the behavior at the
epidemic threshold, $\alpha=1$. We begin with the probability $P_i(t)$
that there are $i$ infected individuals at time $t$. Irrespective of
the infection rate, this distribution function satisfies a closed
evolution equation, and for the critical $\alpha=1$ case, $P_i(t)$
satisfies
\begin{equation}
\label{Pi-eq}
\frac{dP_i}{dt}=(i-1)P_{i-1}+(i+1)P_{i+1}-2iP_i\,.
\end{equation}
The initial condition is $P_i(0)=\delta_{i,1}$. Equations
\eqref{Pi-eq} are closed because in the infinite population limit,
$N\to\infty$, the population of infected individuals does not depend
on the populations of susceptible and recovered individuals.  We
comment that the master equation \eqref{Pi-eq} is a discrete diffusion
equation with a diffusion coefficient equal to the size of the
infected population.

To solve \eqref{Pi-eq} we use the exponential ansatz:
\hbox{$P_i(t)=\Psi(t)\,[\psi(t)]^{i-1}$} for $i>0$, with the initial
conditions $\Psi(0)=1$ and $\psi(0)=0$ to assure the validity of
$P_i(0)=\delta_{i,1}$. This ansatz reduces the infinite set of
equations \eqref{Pi-eq} to two ordinary differential equations,
\hbox{$d\psi/dt=(1-\psi)^2$} and \hbox{$d\Psi/dt=2(\psi-1)\psi$}.
Solving these coupled equations we obtain
\begin{equation}
\label{pi}
P_i(t)=\frac{1}{(1+t)^2}\, \left(\frac{t}{1+t}\right)^{i-1}
\end{equation}
for $i>0$. Thus, the size distribution is purely exponential.

The quantity $P_0(t)$ gives the probability that the infection process
has subsided by time $t$. From \hbox{$dP_0/dt=P_1$} and the initial
condition $P_0(0)=0$, we obtain \hbox{$P_0(t)=t/(1+t)$}. Equivalently,
one can deduce this result from the normalization requirement
$\sum_{i\geq 0} P_i=1$. Note also that the survival probability
$P(t)$, defined as the probability that the infection remains active
at time $t$, is simply $P(t)=1-P_0(t)$, and hence, it is given by
\begin{equation}
\label{survival}
P(t)=\frac{1}{1+t}\,.
\end{equation}
In the long time limit, the survival probability decays algebraically,
$P\simeq t^{-1}$.

Interestingly, the first moment of the distribution $P_i$ is
conserved, $\sum_i iP_i=1$, and in this sense, the competing processes
of infection and recovery balance when $\alpha=1$.  Hence, if we
restrict our attention to active outbreaks, the average number of
infected individuals grows linearly with time, $\langle i\rangle=1+t$,
where $\langle i\rangle =\sum_i i P_i/P(t)$.

In the long-time limit, the distribution $P_i(t)$ has the scaling form
\begin{equation}
\label{pi-scal}
P_i(t)\simeq t^{-2}\,\Phi\left(i\,t^{-1}\right), \qquad
\Phi(\xi)=e^{-\xi}\,.
\end{equation}
This scaling behavior immediately follows from \eqref{pi} in the limit
$i\to\infty$ and $t\to\infty$ with the scaling variable $\xi=i/t$ kept
fixed.

The master equation \eqref{Pi-eq} is closed since the infected
population is decoupled from the other two populations. In contrast,
the recovered population is coupled to the infected population.
Therefore, we must analyze the joint distribution $P_{i,r}(t)$, that
is, the probability there are $i$ infected and $r$ recovered at time
$t$. Of course, the joint distribution function gives a complete
description of the state of the system and for example,
\hbox{$P_i=\sum_{r\geq 0} P_{i,r}$}.  The joint distribution obeys
\begin{equation}
\label{Pir-eq}
\frac{dP_{i,r}}{dt}=(i-1)P_{i-1,r}+(i+1)P_{i+1,r-1}-2iP_{i,r}
\end{equation}
and \hbox{$P_{i,r}(0)=\delta_{i,1}\,\delta_{r,0}$}. In this rate
equation, the first gain term accounts for infection, while the second
gain term represents recovery.  For small $i$ and $r$ one can compute
the joint distribution recursively: $P_{1,0}=e^{-2t}$,
$P_{0,1}=(1-e^{-2t})/2$, $P_{1,1}=e^{-2t}[2t-(1-e^{-2t})]/2$, etc. 
\cite{exp}.

Generally, we use the generating function
\begin{equation}
\label{Puvt}
{\cal P}(u,v)=\sum_{i\geq 0}\sum_{r\geq 0} P_{i,r}\,u^i\,v^r\,.
\end{equation}
By multiplying \eqref{Pir-eq} by $u^i v^r$ and summing over all $i$ and
$r$, we find that the generating function obeys 
\begin{equation}
\label{Puv-eq}
\frac{\partial {\cal P}}{\partial t}
=(u^2-2u+v)\,\frac{\partial {\cal P}}{\partial u}\,.
\end{equation}
The initial condition is \hbox{${\cal P}_0\equiv {\cal P}(u,v)=u$}
where \hbox{${\cal P}_0\equiv {\cal P}\big|_{t=0}$}.

We can transform equation \eqref{Puv-eq} into the wave equation
\begin{equation}
\label{Pw}
\frac{\partial {\cal P}}{\partial t}=\frac{\partial {\cal P}}{\partial w}.
\end{equation}
To establish this equation, we introduce the variable 
\begin{eqnarray}
\label{w-def}
w&=&\int_0^u \frac{du'}{(u')^2-2u'+v} \nonumber \\
 &=&\frac{1}{2\sqrt{1-v}}\,\,
\ln\frac{1+\sqrt{1-v}-u}{1-\sqrt{1-v}-u}\,\,
\frac{1-\sqrt{1-v}}{1+\sqrt{1-v}}\,.
\end{eqnarray}
The solution to equation \eqref{Puv-eq} is simply ${\cal P}_0(w+t)$.
Since ${\cal P}_0=u$, we need to obtain $u$ in terms of the variable
$w$ and then, replace $w$ with $w+t$ to obtain the generating function
explicitly.  {}From \eqref{w-def}, the expression
\begin{eqnarray}
\label{u}
u &=& 1-\sqrt{1-v}\nonumber\\
&-&2\sqrt{1-v}
\left[\frac{1+\sqrt{1-v}}{1-\sqrt{1-v}}e^{2w\sqrt{1-v}}-1\right]^{-1}
\end{eqnarray}
gives ${\cal P}_0=u$ in terms of $v$ and $w$. As a function of the
variables $w$ and $v$, the generating function becomes
\begin{eqnarray*}
{\cal P}(w,v)&=&1-\sqrt{1-v}\\ &-&2\sqrt{1-v}
\left[\frac{1+\sqrt{1-v}}{1-\sqrt{1-v}}\,\,e^{2(w+t)\sqrt{1-v}}-1\right]^{-1}.
\end{eqnarray*}
This result is derived from \eqref{u} by replacing $w$ with $w+t$.  In
terms of the original variables, the generating function is
\begin{equation}
\label{P-solution}
{\cal P}(u,v)=1-\sqrt{1-v}-2\sqrt{1-v}\,\frac{1-\sqrt{1-v}-u}{D-(E-1)u}\,.
\end{equation}
We obtained this result by expressing $\exp\big[2w\sqrt{1-v}\big]$ in
terms of $u$. In writing \eqref{P-solution}, we also used the
shorthand notations
\begin{equation}
\label{ED}
E=e^{2t\sqrt{1-v}}\,,\qquad D=(E+1)\sqrt{1-v}+E-1.
\end{equation}
For example, we can verify that the generating function yields the
size distribution of outbreaks.  In the long-time limit, the last term
on the right-hand-side of \eqref{P-solution} vanishes since the
quantities $D$ and $E$ are both divergent. With the shorthand notation
${\cal P}_\infty\equiv \lim_{t\to\infty} {\cal P}$, we have
\hbox{${\cal P}_\infty(u,v)= 1-\sqrt{1-v}$}.  By substituting
$\alpha=1$ into \eqref{Gz}, we confirm that
$P_{i,r}(t)\to\delta_{i,0}\,G_r$ when $t\to\infty$.

First, we analyze the probability distribution $\Pi_r(t)$. By
definition, $\Pi_r(t)$ is the probability to have $r$ recovered
individuals at time $t$. Of course, $\Pi_r(t)=\sum_{i\geq
0}P_{i,r}(t)$.  The distribution of recovered differs from the
distribution of infected in that in the long-time limit, it approaches
a nonzero value, $\Pi_r\to G_r$, whereas $P_i\to 0$ for all $i>0$. To
study the long-time asymptotic behavior, we write
$\Pi_r(t)=G_r+H_r(t)$. We have $H_r(t)\to 0$ as $t\to\infty$. The
corresponding generating function $H(v)=\sum_{r\geq 0}H_rv^r$ is given
by
\begin{equation}
\label{Hv}
H(v)=\frac{2\sqrt{1-v}}{e^{2t\sqrt{1-v}}+1}\,.
\end{equation}
We obtain this expression from the joint generating function
\eqref{P-solution} by setting $u=1$, $H(v)={\cal P}(1,v)-
P_\infty(1,v)$ where ${\cal P}_\infty(1,v)=1-\sqrt{1-v}$.

According to the scaling behavior \eqref{pi-scal}, the typical size of
the infected population grows linearly with time, $i\sim t$.
Heuristically, $dr/dt \sim i$ since the recovery rate is constant.
Consequently, the typical size of the recovered population is
quadratic, $r\sim t^2$, and hence, we expect the scaling behavior
\begin{equation}
\label{H-scal}
H_r(t)\simeq t^{-3} \varphi(r\,t^{-2}). 
\end{equation}

The Laplace transform of the scaling function $\varphi(\eta)$ follows
immediately from behavior of the generating function \eqref{Hv} in the
limit $v\to 1$. We take the limits $v\to 1$ and $t\to \infty$ while
keeping the variable $b=t^2(1-v)$ fixed. In this limit, we have
$v^r\to e^{-b\eta}$ and the sum over $r$ turns into the integral
\begin{equation*} 
\sum_{r\geq 0} v^r\to t^2\int_0^\infty d\eta\,e^{-b\eta}\,.  
\end{equation*} 
By using this scaling transformation along with Eq.~\eqref{Hv},
we find the Laplace transform of the scaling function
\begin{equation}
\label{varphi-transform}
\int_0^\infty d\eta\, e^{-b\eta}\, \varphi(\eta)=
\frac{2\sqrt{b}}{e^{2\sqrt{b}}+1}\,.  
\end{equation}

The inversion of this Laplace transform through integration in the
complex plane is detailed in the Appendix where we show that the
scaling function $\varphi(\eta)$ is given by
\begin{eqnarray}
\label{varphi-eta}
\varphi(\eta)\!=\!
2\pi^2\sum_{k=0}^\infty \left(k+\tfrac{1}{2}\right)^2\!
e^{-\pi^2\left(k+\frac{1}{2}\right)^2\eta}
-\frac{1}{\sqrt{4\pi\eta^3}}\,.
\end{eqnarray}
We now substitute this expression into the scaling form \eqref{H-scal}
and observe that the algebraic factor $(4\pi)^{-1/2}\eta^{-3/2}$ and
the final distribution $G_r$ cancel each other. Thus, the distribution
$\Pi_r(t)$ has the same scaling form as the distribution $H_r(t)$,
\begin{equation}
\label{pr-scal}
\Pi_r(t)\simeq t^{-3} \phi(r\,t^{-2})\,. 
\end{equation}
The scaling function $\phi(\eta)$ is given by the sum 
\begin{equation}
\label{phieta}
\phi(\eta)=2\pi^2\sum_{k=0}^\infty \left(k+\tfrac{1}{2}\right)^2
e^{-\pi^2\left(k+\frac{1}{2}\right)^2\eta}\,.
\end{equation}
The leading asymptotic behaviors of the scaling function $\phi(\eta)$ are 
\begin{equation}
\label{Phieta-extremal}
\phi(\eta)\simeq 
\begin{cases}
(4\pi)^{-1/2}\eta^{-3/2}\qquad & \eta\to 0;\\
(\pi^2/2)e^{-\pi^2\eta/4} \qquad & \eta\to \infty.
\end{cases}
\end{equation}
The algebraic behavior in the small-$\eta$ limit is consistent with
the power-law tail \eqref{Gn-large}. To obtain this behavior, we
simply convert the sum in \eqref{phieta} into an integral. We note
that with the scaling form \eqref{pr-scal} and the small-$\eta$
divergence, the quantity $\sum_r \Pi_r$ is indeed finite.  Also, the
first term in the series yields the exponential behavior in the
large-$\eta$ limit. Hence, both the distribution of recovered and the
distribution of infected have exponential tails.

We now analyze the joint distribution $P_{i,r}(t)$. We restrict our
attention to active epidemics that correspond to $i>0$. Thus, we focus
on the following component of the joint generating function
\begin{eqnarray}
\label{fuv-def}
{\cal F}(u,v)=\sum_{i\geq 1} \sum_{r\geq 0} P_{i,r} u^i v^r.
\end{eqnarray}
This component follows directly from the generating function, ${\cal
  F}(u,v)={\cal P}(u,v)-{\cal P}(0,v)$, and by using the explicit
solution \eqref{P-solution}, we obtain
\begin{eqnarray}
\label{fuv}
{\cal F}(u,v)=\frac{4E u(1-v)}{D[D-(E-1)u]}.
\end{eqnarray}

The scaling behaviors \eqref{pi-scal} and \eqref{pr-scal} for the
single-population distributions imply that the joint distribution
function adheres to the scaling form \eqref{Pir-scal} stated in the
introduction.  Since the survival probability $P(t)=\sum_{i\geq
1}\sum_r P_{i,r}(t)$ decays with time according to \eqref{survival},
the scaling function is normalized $\iint d\eta\, d\xi\,
\Phi(\xi,\eta)=1$.

The above analysis suggests use of the joint Laplace transform
\begin{equation}
\label{fab-def}
f(a,b)=\int_0^\infty \int_0^\infty d\xi\, d\eta\, 
e^{-a\xi-b\eta}\,\Phi(\xi,\eta),
\end{equation}
that is merely the continuous counterpart of the generating
function. To obtain $f(a,b)$ from ${\cal F}(u,v)$ given in
\eqref{fuv}, we take the limits $t\to\infty$, $u\to 1$, and $v\to 1$
while keeping the variables
\begin{equation*}
a=(1-u)\,t\quad {\rm and}\quad b=(1-v)\,t^2
\end{equation*}
fixed. By taking these limits, we observe that the right-hand side of
\eqref{fuv-def} becomes $t^{-1}f(a,b)$, and find the joint Laplace
transform in  explicit form
\begin{equation}
\label{fab}
f(a,b)=
\left(\frac{\sqrt{b}}{\sinh\sqrt{b}}\right)^2
\frac{1}{a+\sqrt{b}\coth\sqrt{b}}.
\end{equation}
We stress that the quantity $f(a,b)$ describes only active infection
processes.  One can verify that $f(0,0)=1$. Moreover, in the $b\to 0$
limit, we have \hbox{$f(a,0)=1/(1+a)$}, for which the inverse Laplace
transform is immediate, $\int d\eta\,
\Phi(\xi,\eta)=e^{-\xi}$. Indeed, we recover the scaling function
$\Phi(\xi)$ in Eq.~\eqref{pi-scal}.

Since the inverse Laplace transform of $f(a,b)$ with respect to the
variable $b$ is immediate, we have
\begin{equation}
\label{Phi-xib}
\int d\eta\, e^{-b\eta}\Phi(\xi,\eta)=
\left(\frac{\sqrt{b}}{\sinh\sqrt{b}}\right)^2e^{-\xi\sqrt{b}\coth\sqrt{b}}.
\end{equation}
The Appendix outlines how to invert this Laplace transform to obtain
the leading asymptotic behaviors for large-$\eta$ and small-$\eta$,
\begin{equation}
\label{Fxieta-extremal}
F(\xi,\eta)\simeq 
\begin{cases}
\frac{4\,(1+\xi/2)^3}{\pi^{1/2}\,\eta^{7/2}}
\,\,\exp\left[-\frac{(1+\xi/2)^2}{\eta}\right] & \eta \to 0, \\
\frac{\pi^2\,2^{1/4}}{\xi^{3/4}}\,\,\exp\left[-\pi^2\eta+\pi\sqrt{8\xi\eta}-\xi\right] & \eta \to \infty.
\end{cases}
\end{equation}
These limiting behaviors apply for a fixed value of $\xi$.

The function $f(a,b)$ captures all moments of the scaling function
$\Phi(\xi,\eta)$. For example, by expanding equation \eqref{fab} as a
Taylor series in the conjugate variables $a$ and $b$,
$f(a,b)=1-a-\tfrac{2}{3}b+ab+\cdots$, and comparing with the
definition \eqref{fab-def}, we obtain the lowest-order moments
$\langle \xi\rangle =1$, $\langle \eta\rangle =\tfrac{2}{3}$, and
$\langle \xi\eta\rangle =1$.  In particular,
\begin{equation}
\langle \xi\eta\rangle >\langle \xi\rangle \langle \eta\rangle,
\end{equation}
so there is positive correlation between the size of the two
populations. Intuitively, we expect that long-lasting epidemic
outbreaks involve large numbers of infected and recovered, while the
opposite is true for short-lived outbreaks.

For completeness, we mention that the above analysis can be repeated
for inactive outbreaks. Starting with ${\cal P}(0,v)$ given in
\eqref{P-solution} and following the steps leading to \eqref{pr-scal},
we find that the distribution $P_{0,r}(t)$ that an epidemic outbreak
has ended by time $t$ and that the size of the outbreak equals $r$,
has the scaling form \hbox{$P_{0,r}(t)\simeq
  t^{-3}\tilde\phi\big(r\,t^{-2}\big)$}. The scaling function
$\tilde\phi(\eta)$ resembles $\phi(\eta)$ given in \eqref{phieta}
\begin{equation}
\tilde\phi(\eta)=2\pi^2\sum_{k=1}^\infty k^2\,e^{-\pi^2 k^2\eta}.
\end{equation}
The leading asymptotic behavior in the small-$\eta$ limit is identical
to that in \eqref{Phieta-extremal}, and again, there is exponential
decay, albeit with twice the coefficient,
\hbox{$\tilde\phi(\eta)\simeq 2\pi^2 \exp(-\pi^2\eta)$} in the
large-$\eta$ limit. As expected, inactive outbreaks dominate the size
distribution of the recovered population at large times.

\section{Finite Populations}

The zeroth and first-order moments of the distribution $G_r$ given in
equations \eqref{G-finite} and \eqref{nav} show that the size of the
epidemic outbreak is microscopic below the epidemic threshold,
$\alpha<1$, but macroscopic above the threshold, $\alpha>1$. These
results hold for infinite populations, yet real-world epidemic
outbreaks involve finite populations, and there is practical need for
understanding how basic characteristics such as the average size and
the average duration of the epidemic depend on the size of the
population, $N$.

Finite-size effects are most pronounced in the vicinity of the
epidemic threshold. Let us consider large but finite populations,
$N\gg 1$, and let us consider a scenario in which the susceptible
population has been depleted by $n$, that is, $s=N-n$. From the very
definition of the infection process \eqref{infection}, we determine
that the infection rate is reduced, $\alpha \to \alpha_*(N)$, with
\begin{equation}
\label{alpha-n}
\alpha_*(N)=\alpha \left(1-\frac{n}{N}\right),
\end{equation}
because the total population is finite.  Clearly, as the large
``reservoir'' of susceptible individuals shrinks, the infection
process weakens.

\begin{figure}[t]
 \includegraphics*[width=0.45\textwidth]{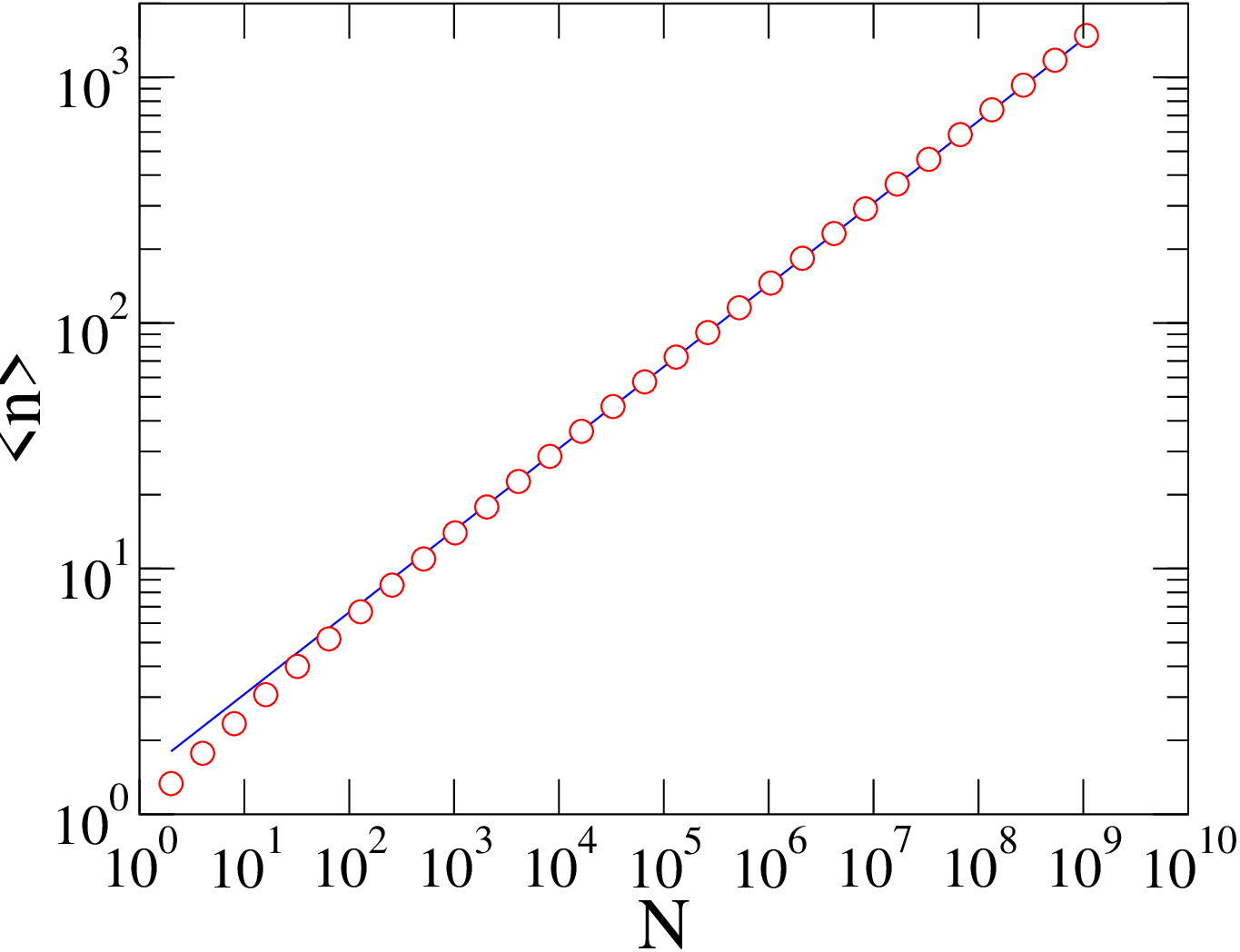}
 \caption{The average outbreak size versus the population size for the
 SIR infection process at the epidemic threshold ($\alpha=1$). Shown
 are Monte Carlo simulation results representing an average over
 $10^9$ independent realizations of the infection process (circles). A
 line of slope $1/3$ is also shown as a reference.}
\label{fig-nav}
\end{figure}

Let us now consider the average size of the outbreak for a threshold
epidemic ($\alpha=1$). We assume that there is a maximal outbreak size
$M$, and that the outbreak can not exceed this size due to depletion
in the number of susceptible individuals.  On the one hand, equations
\eqref{alpha-n} and \eqref{nav} suggest that $\langle n\rangle \sim
N/M$. On the other hand, the algebraic distribution \eqref{Gn-large}
gives a second estimate $\langle n\rangle \sim \sum_{n\leq M} n^{-1/2}
\sim M^{1/2}$ for the average size of the outbreak.  Equating these
two estimates, $N/M\sim M^{1/2}$, we conclude that $M\sim N^{2/3}$ as
stated in \eqref{MT}. While a naive interpretation of the effective
infection rate \eqref{alpha-n} would suggest that the cutoff is
proportional to the total population, we find that threshold epidemics
have a substantially smaller upper bound.

\begin{figure}[t]
 \includegraphics*[width=0.45\textwidth]{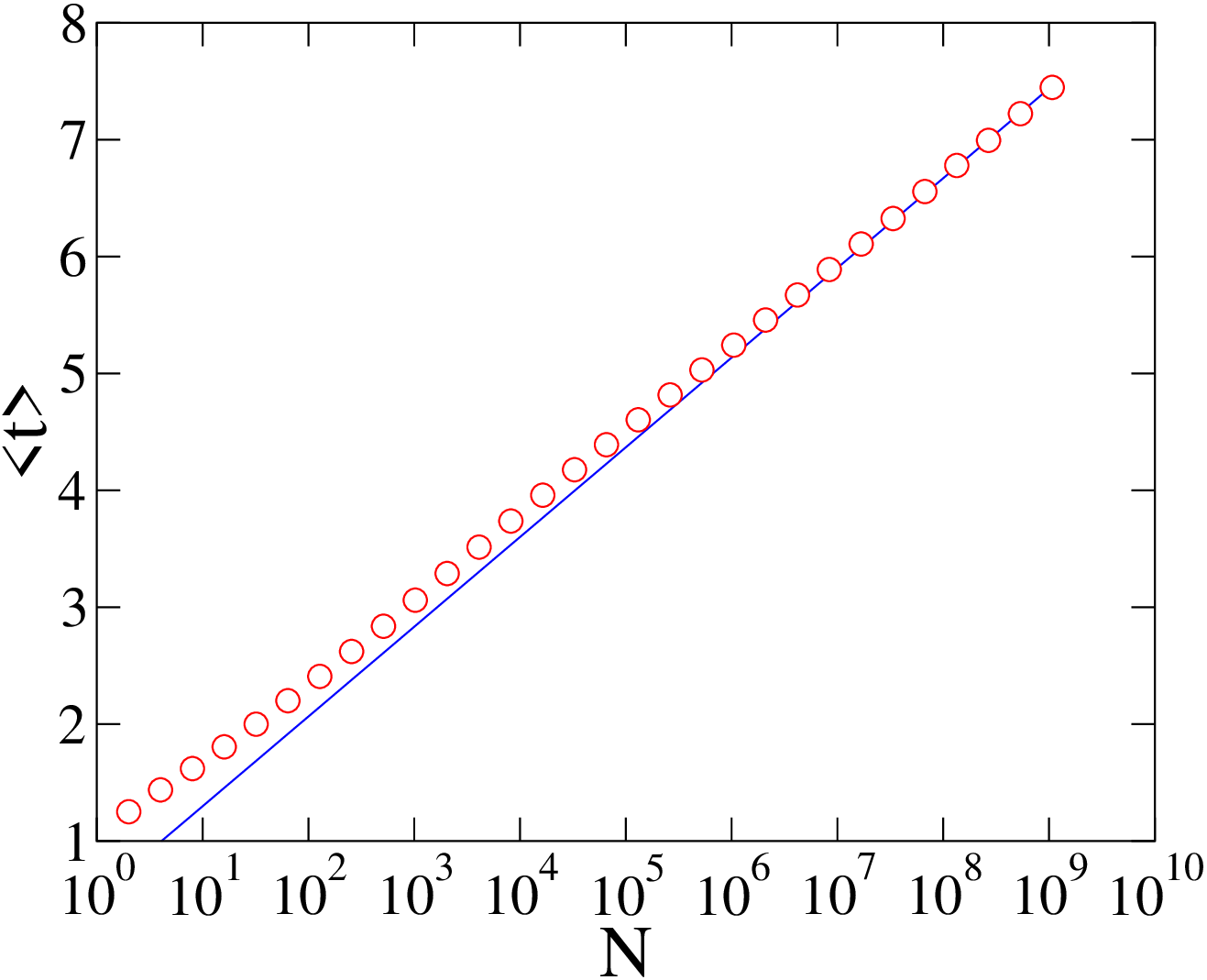}
\caption{The average outbreak duration at the epidemic threshold
  versus the population size. Simulation results, obtained from an
  average over $10^9$ realizations (circles) are compared with the
  reference line $(1/3)\ln N+C$. }
 \label{fig-tav}
\end{figure}

Using $\langle n\rangle \sim N/M$ and $M\sim N^{2/3}$ we see 
that the average size of a threshold epidemic grows as  
\begin{equation}
\label{nav-n}
\langle n\rangle \sim N^{1/3}. 
\end{equation}
This interesting scaling law was established by Martin-L\"{o}f
\cite{ML} and it has been confirmed in several recent studies
\cite{bk,KS,ML2,Dutch,KS1}.  Figure \ref{fig-nav} shows a numerical
verification of this behavior.  For finite populations, this scaling
law holds in the neighborhood (often termed critical or scaling
windows) of the epidemic threshold. To estimate the size of this
neighborhood, we simply compare \eqref{nav-n} with \eqref{nav}. The
size of the outbreak grows sub-linearly in the population size as long
as
\begin{equation}
\label{neighborhood}
|1-\alpha|\sim N^{-1/3}.
\end{equation}
Of course, this neighborhood shrinks as the size of the population
grows. Yet, for moderate populations, the size of this neighborhood is
considerable.

The scaling law \eqref{nav-n} indicates that for finite populations,
there are actually three regimes of behavior. Well below the epidemic
threshold, a finite {\em number} of individuals becomes infected. Well
above the threshold, a finite {\em fraction} of the population becomes
infected. In the vicinity of the threshold, the size of the outbreak
grows as the $1/3$ power of the population size,
\begin{equation}
\label{nav-three}
\langle n\rangle\sim
\begin{cases}
{\cal O}(1) \qquad      & \zeta\to \infty ,\cr
N^{1/3}\,Y(\zeta)\qquad    & \zeta={\cal O}(1),\cr
N      \qquad    & \zeta\to -\infty ,\cr
\end{cases}
\end{equation}
with the scaling variable $\zeta=N^{1/3}(1-\alpha)$. A finite value of
$\zeta$ indicates that the infection rate is in the neighborhood of
the threshold. (The scaling function $Y(\zeta)$ can be extracted from 
Refs.~\cite{ML,KS,ML2,Dutch}.)

The duration of a threshold epidemic also adheres to a finite-size
scaling law. Since the size of the recovered population grows
quadratically with time, $r\sim t^2$, as implied for example by the
scaling behavior \eqref{pi-scal}, we conclude that there is a maximal
time scale $T$, and that the duration of a near-threshold outbreak may
not exceed this scale. We can estimate the time scale $T\sim N^{1/3}$
stated in \eqref{MT}, from the heuristic argument \hbox{$T^2\sim
M$}. The average duration of the infection process follows from the
survival probability, $\langle t\rangle = -\int_0^T dt \, t\,dP/dt$,
and using $P\sim t^{-1}$, we find that the average duration grows
logarithmically with time (Fig.~\ref{fig-tav}),
\begin{equation}
\label{tav-n} 
\langle t\rangle\simeq\frac{1}{3}\,\ln N\,.
\end{equation}
Logarithmic growth, albeit with a unit prefactor, was predicted by
Ridler-Rowe \cite{rr}.

\begin{figure}[t]
 \includegraphics*[width=0.45\textwidth]{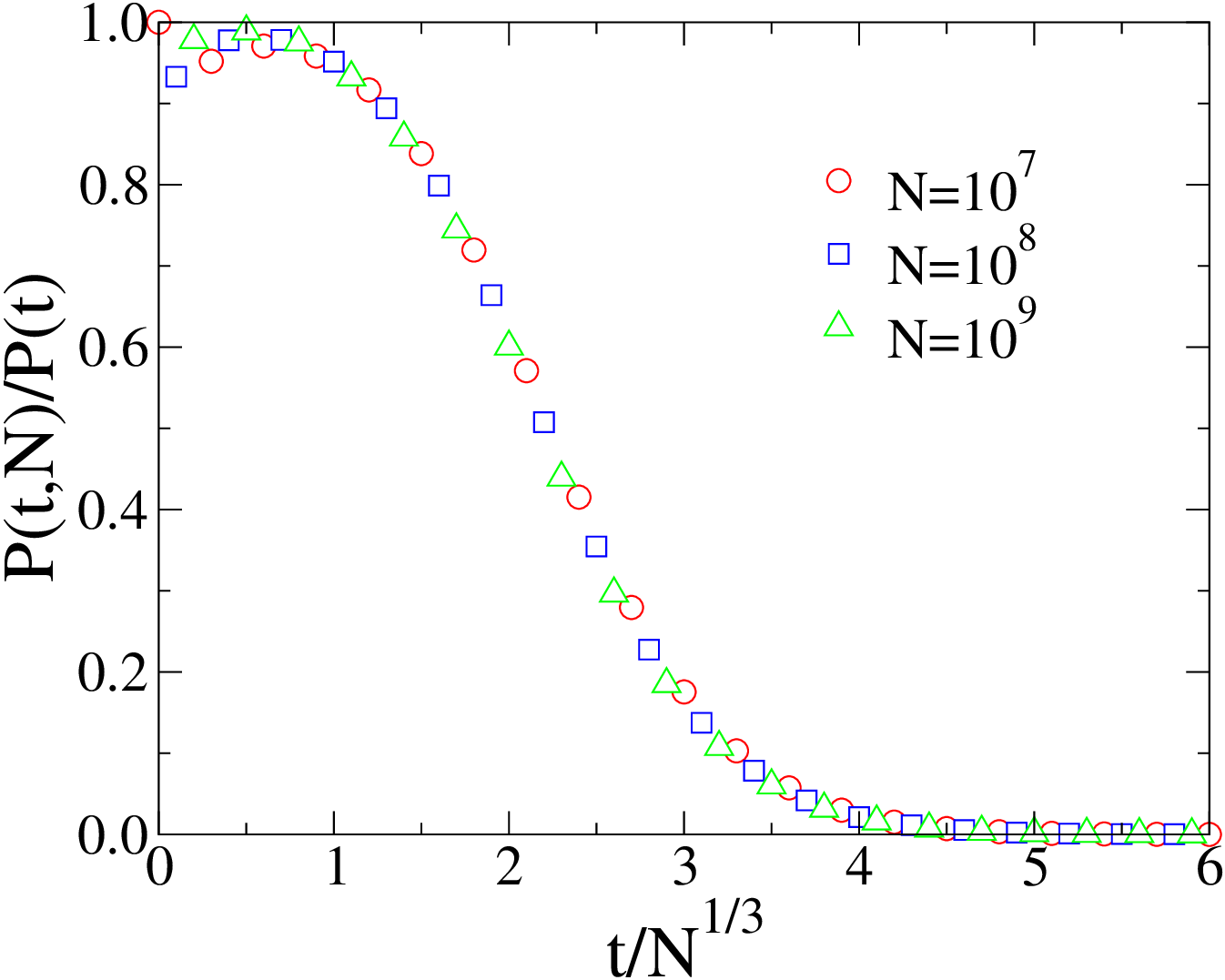}
\caption{The survival probability at the epidemic threshold. Shown is
the normalized survival probability $P(t,N)/P(t)$ versus the
normalized duration time $t/N^{1/3}$. The data corresponds to an
average over $10^8$ realizations.}
\label{fig-scal}
\end{figure}

The scales $M\sim N^{2/3}$ and $T\sim N^{1/3}$ fully characterize the
size and duration of the infection process when \hbox{$\alpha=1$}. For
example, the survival probability $P(t,N)$ to have at least one
infected at time $t$ obeys $P(t,N)/P(t)\to {\cal
S}\left(t/N^{1/3}\right)$ (Fig.~\ref{fig-scal}). However, the
simulations show that the convergence to this scaling form is not
uniform --- it is slow for short durations but fast at large
durations.

We performed numerical simulations to check the scaling predictions
for the case $\alpha=1$. Our numerical method is merely a Monte Carlo
procedure to solve the master equation \eqref{Pir-eq}. We conveniently
keep track of two populations, $s$ and $i$, from which the overall
infection rate and recovery rate are respectively $R_i=si/N$ and
$R_r=i$.  The probabilities $P_i=R_i/(R_i+R_r)$ and
$P_r=R_r/(R_i+R_r)$ that the following step involves infection or
recovery are calculated and then, the populations are updated
accordingly,
\begin{equation}
(s,i)\to 
\begin{cases}
(s-1,i+1) & {\rm with\ probability\ } P_i, \\
(s,i-1) & {\rm with\ probability\ } P_r .
\end{cases}
\end{equation}
The infection process ends when $i=0$ for the first time. Time is
updated by the inverse of the total rate \hbox{$t\to t+1/(R_i+R_r)$}.
With this straightforward numerical procedure, we can simulate as many
as $10^9$ independent runs in populations as large as $N=10^9$. As
shown in figures \ref{fig-nav} and \ref{fig-tav}, there is excellent
agreement between the numerical results and the theoretical
predictions for the average size \eqref{nav-n} and the average
duration \eqref{tav-n}.

\section{Discussion}

We investigated various statistical properties of the SIR infection
process at a threshold infection rate. We analyzed the rate equations
for the two-population distribution function that characterizes the
probability that the system has a specified number of infected and
recovered individuals. Our analysis yields scaling behavior in the
asymptotic long-time limit and gives extremal properties of the joint
distribution function as well as the associated single-population
distributions. We used these infinite population results to justify
scaling laws for finite populations.

Outbreaks in the vicinity of the epidemic threshold have a distinct
size, characterized by non-trivial power-law dependence on the total
population size.  This scaling behavior applies in the vicinity of the
epidemic threshold. The size of this neighborhood, $N^{-1/3}$, is
larger than the canonical value $N^{-1/2}$ expected from the
traditional large-size expansions or from the deterministic
description \cite{nvk}.  Therefore, statistical fluctuations and
finite population effects are pronounced and may be subtle near the
epidemic threshold.  We note the an identical scaling arises near the
percolation point for Erd\H{o}s-R\'enyi random graphs
\cite{bbckw,bk05}. Additional connections between SIR infection
processes and random graphs were reported recently \cite{Dutch}.

The scaling laws for the time dependence and the size dependence are
useful. For example, the scaling laws for the critical kinetics can be
used to find the epidemic threshold in numerical simulations of
infection processes on complex networks for which the threshold is not
known analytically. Furthermore, the number of coupled ordinary
equations needed to compute the joint distribution $P_{i,r}$
numerically is in principle quadratic with $N$. However, the scaling
laws $i\sim N^{1/3}$ and $r \sim N^{2/3}$ imply that the relevant
number of equations is much smaller, being proportional to $N$.

For large but finite populations, we understand the basic scaling
laws, but much less is known about finite-size scaling functions. The
only exception is the known scaling function that gives the size
distribution of the outbreaks at the epidemic threshold
\cite{ML,KS}. The analytic determination of the scaling function
characterizing the duration of outbreaks near the epidemic threshold
is a challenging problem because there is no closed equation for the
total duration of the outbreak \cite{book}.

Finally, we mention that in this study we assumed that all individuals
can interact. In most applications, the spatial
\cite{pg,sa,ad1,hin,tibor,dickman,jdmII,wss,ziff} or
network \cite{pv,ml,mejn} structure of the infected domain play an
important role. Finding the corresponding scaling functions for the
time-dependent behavior at the critical point or for the finite-size
scaling are also challenging open problems.

\acknowledgments
We are grateful to Aric Hagberg for initial collaboration and thank
Tibor Antal and David Kessler for useful correspondence.  We also
acknowledge support for this research by DOE grant DE-AC52-06NA25396.

\appendix

\section{Inverse Laplace Transforms}
\label{Ap-Lap}

From Eq.~\eqref{varphi-transform}, the scaling function
$\varphi(\eta)$ equals the inverse Laplace transform
\begin{equation}
\label{inverse}
\varphi(\eta)=\frac{1}{2\pi i} \int_{\gamma-i\infty}^{\gamma+i\infty}
db\,e^{b\eta}\, \frac{2\sqrt{b}}{e^{2\sqrt{b}}+1}
\end{equation}
where the integration contour is a line parallel to the imaginary axis
in the complex $b$ plane and $\gamma>0$ so that all singularities are
to the left of the integration contour.  The integrand has simple
poles at $b_k=-\pi^2\left(k+\frac{1}{2}\right)^2$, $k=0, 1, 2, \ldots$
and a branch point at the origin, and we select a branch cut along the
negative real axis so that it does not cross the path of integration.
Furthermore, we pick a closed Bromwich contour formed by the contour
followed by a quarter of a circle of infinitely large radius, followed
by the top of the branch cut with infinitesimal half-circles around
each pole and then encircling the origin and proceeding similarly
along the bottom of the branch cut, followed by a quarter of a circle.
By the Cauchy theorem, the integral along this closed contour
vanishes. The integrals over the circles vanish and the integrals over
the branch cut can be computed to yield 
\begin{eqnarray*}
\varphi(\eta)\!=\!2\pi^2\sum_{k=0}^\infty \!\!\left(k+\tfrac{1}{2}\right)^2\!
e^{-\pi^2\left(k+\tfrac{1}{2}\right)^2\eta}-\frac{1}{\pi} \int_0^{\infty}\!\! dc\,e^{-c\eta}\,\sqrt{c}
\end{eqnarray*}
where the sum is proportional to the residues of poles at
$b_k=-\pi^2\left(k+\frac{1}{2}\right)^2$.  Since the integral on the
right-hand side equals $(4\pi)^{-1/2}\eta^{-3/2}$, we arrive at
\eqref{varphi-eta}.

The inverse Laplace transform of \eqref{Phi-xib} with respect to $b$
gives the joint scaling function 
\begin{equation*} 
F(\xi,\eta)=\frac{1}{2\pi i}
\int_{\gamma-i\infty}^{\gamma+i\infty} db\,e^{b\eta}\,
\left[\frac{\sqrt{b}}{\sinh \sqrt{b}}\right]^2
e^{-\xi\,\sqrt{b}\,\coth \sqrt{b}}.
\end{equation*} 
To obtain the leading asymptotic behavior in the small-$\eta$ limit,
given in \eqref{Fxieta-extremal}, we simply evaluate the large-$b$
behavior. In the opposite limit, $\eta\to \infty$, the asymptotic
behavior follows from the singularity of the Laplace transform in the
complex $b$ plane that is closest to the origin.  The Laplace
transform has singularities at $b_k=-\pi^2k^2$, $k=1, 2, \ldots$, so
we focus on the singularity at $b_1=-\pi^2$. For a fixed $\xi$, the
scaling function appears to decay exponentially, $F(\xi,\eta)\sim
e^{-\pi^2\eta}$ as $\eta\to \infty$.  To establish the complete
asymptotic behavior we recall that a subdominant power-law prefactor,
$F(\xi,\eta)\sim \eta^m\,e^{-\pi^2\eta}$, implies that the Laplace
transform has the algebraic singularity $(\pi^2+b)^{-m-1}$ as $b\to
-\pi^2$. In fact, the Laplace transform has the essential singularity
(we set $b=-\pi^2(1-\epsilon)^2$, so that the limit $b\to -\pi^2$ is
equivalent to the $\epsilon\to 0$ limit)
\begin{equation}
\label{essential}
\epsilon^{-2}\,\exp\left[({\epsilon}^{-1}-1)\xi\right]. 
\end{equation}
An essential singularity of the type $e^{\xi/\epsilon}$ corresponds to
a subdominant prefactor that is a stretched exponential,
$e^{\pi\sqrt{8\xi\eta}}$.  An additional power-law factor times a
numerical factor allow to match the singularity (\ref{essential}). The
resulting large-$\eta$ asymptotic behavior is given in
\eqref{Fxieta-extremal}.


\begin{thebibliography}{99}
\bibitem{exp}
    The distribution describing the populations with no recovered is easy to compute;
    it turns out to be exponential, 
    $P_{i,0}=e^{-2t}\big[(1-e^{-2t})/2\big]^{i-1}$.
\end{thebibliography}
\end{document}